\documentclass[10pt,prl,twocolumn,showkeys,showpacs,floatfix]{revtex4}

\usepackage{amsmath}
\usepackage{amssymb}
\usepackage{graphicx}
\usepackage{epstopdf}
\usepackage{dcolumn}

\usepackage{natbib}
\begin{document}

\title{Effect of non-linear interface kinetics on coarsening phenomena}

\author{M. \surname Upmanyu}
\email{mupmanyu@mines.edu}
\affiliation{Group for Simulation and Theory of Atomic-scale Material Phenomena (stAMP),
Engineering Division, Materials Science Program,
Bioengineering and Life Sciences Program, Colorado School of Mines, Golden, CO 80401}
\author{P. A. \surname Martin}
\affiliation{Department of Mathematical and Computer Sciences, Colorado School of Mines, Golden, CO 80401}
\author{A. D. \surname Rollett}
\affiliation{Department of Materials Science and Engineering, Carnegie Mellon University, Pittsburgh, PA 15213}
                
\begin{abstract}
Coarsening kinetics is usually described using a linear gradient approximation for the underlying interface migration (IM) rates, wherein the migration fluxes at the interfaces vary linearly with the driving force. Recent experimental studies have shown that coarsening of nanocrystalline interface microstructures is unexpectedly stable compared to conventional parabolic coarsening kinetics. Here, we show that during early stage coarsening of these microstructures, IM rates can develop a non-linear dependence on  the driving force, the mean interface curvature. We derive the modified mean field law for coarsening kinetics. Molecular dynamics simulations of individual grain boundaries reveal a sub-linear curvature dependence of IM rates, suggesting an intrinsic origin for the slow coarsening kinetics observed in polycrystalline metals.
\end{abstract}

\maketitle
Material properties of most inorganic polycrystals depend on the underlying interfacial microstructure, in particular the grain size. The positive free energy of the interfaces provides a universal driving force for grain coarsening such that a curved interface moves towards its center of curvature in order to decrease the total system energy. A quantitative understanding of the curvature dependence of the interface migration (IM) rate is central to much of material processing as it  determines the overall microstructure coarsening kinetics. 

In systems where interface motion is activated, the atomic-exchange across the interface determines the IM rate, $v$. Absolute reaction rate theory based on such atomic hopping events yields the dependence of the IM rate on the driving force $p$~\cite{gbm:Turnbull:1951},
\begin{equation}
\label{eq:subLinearComplete}
v = bN\nu_o e^{-\beta \Delta G_A} \left(1 - e^{-\beta\Delta g}\right),
\end{equation}
where $b$ is the interface displacement per event, $N$ is the number of event sites, $\nu_o$ is a jump frequency characteristic of the underlying lattice and $\Delta G_A = \Delta Q_A - T\Delta S_A$ is the activation barrier for each hopping event. System free energy change per event $\Delta g$ is the energy provided by the driving pressure $p$ to effect the atom-exchange, i.e. $\Delta g = p \omega_m$, where $\omega_m$ is the activation volume associated with each event. At high temperatures, $\beta p\omega_m \ll 1$ 
and Eq.~\ref{eq:subLinearComplete} can be linearized to express the IM rate in terms of the driving force, $v  \approx M p$. The constant of proportionality $M$ is the interface mobility, which is predominantly Arrhenius with temperature
\begin{equation}
\label{eq:linearRelations}
M = M_o e^{-\beta\Delta Q_A}\;\text{and}\;M_o = \beta bN\nu_o\omega_m\;e^{\beta\Delta S_A}.
\end{equation}

During coarsening, the capillary driving force on each interface segment is the product of the interface stiffness $\Gamma$ and its mean curvature $\kappa$, or the weighted mean curvature $\kappa_\gamma = \Gamma\kappa$~\cite{cg:Taylor:1992b}. The IM rate now increases with the mean interface curvature~\cite{cg:Taylor:1992, book:SuttonBalluffi:1995, gbm:Upmanyu:1998a}. The curvature dependence implies that material systems with ultrafine/nanocrystalline (nc) grain sizes possess an inherently high driving force for coarsening. This is often an unwanted outcome during thermal annealing, as the qualitatively superior thermomechanical and transport properties of these microstructures are offset by their instability with respect to coarsening. 

However, a poorly understood feature of nanocrystalline microstructures is that the coarsening is anomalously slower than expected, almost linear in time, before it transitions to the conventional parabolic growth~\cite{nano:Bonetti:1999, nano:Krill:2001}. This behavior has been attributed to grain size dependent extrinsic effects on interface motion, such as enhanced vacancy and triple junction drag~\cite{nano:Krill:2001, gbm:Estrin:2000, gbm:Upmanyu:1998b, tjm:Upmanyu:1999}, solute segregation~\cite{imdrag:Michels:1999} and particle incidence. In this article, we present an alternative framework for coarsening behavior based on intrinsically non-linear IM rates, and discuss the implications of this behavior for the stability of nanocrystalline microstructures. 


The rationale is the observation that capillary driving pressures during early stages of coarsening in nanocrystalline microstructures are large enough such that the linearized rate theory begins to break down. This is evident from Table~\ref{tab:criticalGrainSizes}, a list of the critical driving forces $p_{cr}$ in various polycrystalline metals at which $\beta p_{cr}\omega_m = 0.1$. The definition corresponds to a processing temperature $T=0.7\,T_m$ and assumes single-atom hops across the interface, $\omega_m=\Omega$ ($T_m$ is the bulk melting point and $\Omega$ is the atomic volume).  Since the atomic-scale mechanism can in general involve more than one atom, as in correlated or military atom transfers across the interface~\cite{book:SuttonBalluffi:1995, gbm:Upmanyu:2004}, $\omega_m\ge\Omega$ and the reported values of $p_{cr}$ are upper bounds. 
\begin{table}
[thdp]
\begin{center}
\begin{tabular}{|c|c|c|c|}
\hline
{\bf Metal system} & {\bf Volume} $\omega_m$ $\rm {(\AA)}^3$ & $p_{cr}$ {\bf (MPa)}  & $\bar{R}_{cr}$ {\bf (nm)}\\ 
\hline
Aluminum & 16.6 &  54.4 & 25.1\\
\hline
Copper & 11.8 & 87.7 & 51.5\\
\hline
Nickel & 10.9& 153.2 & 18.3\\
\hline
Lead & 30.3 &19.1 &146.4\\
\hline 
Gold & 17.1 &76.1 & 36.8\\
\hline 
Silver & 16.9 & 70.8 & 39.6\\
\hline
\end{tabular}
\end{center}
\caption{Critical driving forces for interface migration and critical grain radii in polycrystalline metals at which linearized reaction rate theory breaks down. The activation volume for IM is assumed to be the atomic volume, $\omega_m=\Omega$.
\label{tab:criticalGrainSizes}}
\end{table}

The critical driving force can also be used to define a critical grain size, $\bar{R}_{cr}$, as the average grain size in polycrystals is a measure of the capillary driving force per interface segment. The mean field relation takes the form $\kappa_\gamma=\Gamma B/\bar{R}$, where the topological parameter $B$ captures the effect of the interface network~\cite{gg:Mullins:1956}. Table 1 also lists the lower bound for the associated critical grain sizes below which the non-linearities become important. The interface stiffness is assumed to $\Gamma=0.5$\,J/m$^2$, while the topological parameter is based on Potts model grain growth simulation-based comparison between shrinking of an embedded sphere ($B=4\pi$) and a 3D polycrystal, $B\sim2.8$~[ADR, to be published]. These critical grain sizes lie well within the range of those found in nanocrystalline microstructures in these metals, emphasizing the role of non-linear interface migration kinetics in determining the overall coarsening kinetics and therefore their thermo-mechanical stability.

\begin{figure} [htb]
\includegraphics[width=\columnwidth]{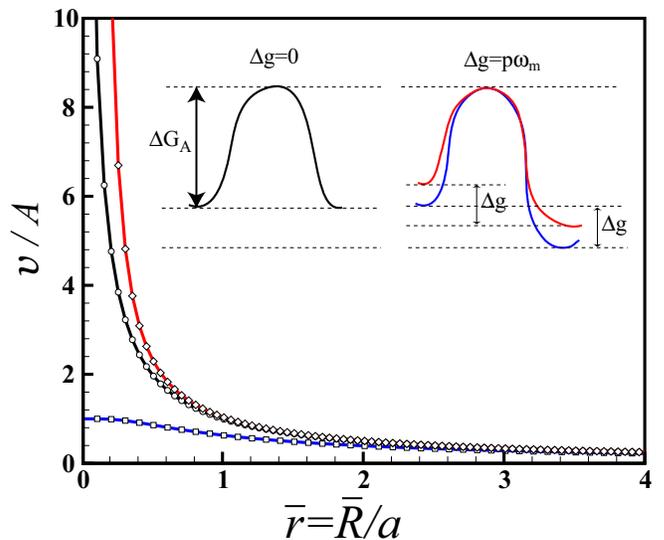}
\caption{\label{fig:subsuper-linearMigrationRateVsRbya} (color online). Normalized migration rate as a function of reduced grain size $\bar{r}=\bar{R}/a$, for linear (circles), sub-linear (Eq.~\ref{eq:GGEquationComplete} with $\xi=0$, squares) and super-linear (Eq.~\ref{eq:GGEquationComplete} with $\xi=0.5$, diamonds) driving force dependence of IM rates. (inset) Schematic of the energy landscapes for in the absence and presence of a driving force, for the two values of $\xi$.}
\end{figure}

The extension of the reaction rate framework to non-linear IM is not straightforward. The activation barrier is now comparable to the driving force, the bias in the energy landscape (inset, Fig~\ref{fig:subsuper-linearMigrationRateVsRbya}). As a result, the nature of the bias becomes important. While several minimum energy paths (MEP) are possible depending on the effect of the driving force on the initial, activated and final states,  we limit ourselves to two extreme scenarios : a) the forward activation barrier is unaltered while the final state is lowered by the applied bias, or b) the bias is distributed equally between the initial and final states such that the forward barrier is altered. In general, the energy of the activated state in the resultant MEP is also changed due to the driving force. Making the simplifying assumption that this energy change is small, Eq.~\ref{eq:subLinearComplete} can be generalized as
\begin{equation}
\label{eq:subSuperLinearComplete}
v= A \left[e^{\xi \beta p\omega_m} - e^{(\xi-1)\beta p\omega_m}\right],
\end{equation}
where $A=\beta M/\omega_m$~\footnote{Change in the energy of the activated state can also be absorbed in a similar form. If this increase is $\zeta p\omega_m$, the pre-factor $A$ for the IM rate is scaled by $e^{-\zeta \beta p\omega_m}$.
}. 
Equation~\ref{eq:subSuperLinearComplete} is similar in form to the Butler-Volmer equation in electrokinetics for transfer of charged species~\cite{book:BardFaulkner:1980}. 

The parameter $\xi$ captures the combined effect of the degree of symmetry in the energy landscape imposed by the driving force, and the change in energy of the activated state. The two extreme cases correspond to $\xi=0$ and $\xi=0.5$. The $\xi=0$ scenario can arise if the driving force is distributed asymmetrically such that its effect is restricted to the final state. The scenarios $0<\xi\le0.5$ result in a decrease in effective forward barrier with increasing driving force. For $\xi=0.5$, the bias is distributed symmetrically about initial and final states such that the forward activation barrier is $\Delta \tilde{G}_A=\Delta G_A - (\Delta g)/2$ (see inset Fig.~\ref{fig:subsuper-linearMigrationRateVsRbya}). Changes in the activated state can be readily absorbed into an equation of this form. For example, if the MEP for the $\xi=0.5$ scenario is modified such that the activated state also increases by an equal amount $(\Delta g/2)$, we recover the $\xi=0$ scenario.

Limiting our analysis to the two scenarios, the IM rates $v(\xi)$ can be expressed as
\[
v(0) = A (1 - e^{-\beta p\omega_m})\;\textrm{and}\; v(0.5) = 2A\sinh(-\beta p\omega_m/2).
\]
In the limit $\beta p\omega_m\gg1$, $v$ becomes independent of $\xi$, justifying the linearization (Eq.~\ref{eq:subLinearComplete}) at small driving forces. In the mean field limit, the IM rate depends on the grain size, $v=d\bar{R}/dt$ and we arrive at the relation between the form of the energy landscape and the grain size evolution,
\begin{align}
\label{eq:GGEquationComplete}
a\frac{d\bar{r}}{dt} &= A \left[e^{\xi \beta p\omega_m}- e^{(\xi-1)\beta p\omega_m}\right].
\end{align}
Here $\bar{r}=\bar{R}/a$ is the dimensionless grain size. The normalization factor $a=B\Gamma\beta\omega_m$ is a fundamental microstructural length scale that is related to the critical grain size, $\bar{R}_{cr}=0.1 a$. Figure~\ref{fig:subsuper-linearMigrationRateVsRbya} shows the IM rates predicted by Eq.~\ref{eq:GGEquationComplete} as a function of reduced grain size for the two scenarios. The linear approximation is also plotted for comparison. At small grain sizes such that $\bar{r}\rightarrow1$, the variation of IM rate is significantly non-linear with the driving force. The linearized relation is an overestimate (sub-linear IM rate) or an underestimate (super-linear IM rate) depending on the value of $\xi$, underscoring the role of the nature of the bias in the energy landscape. 

The non-linearities in IM rates will modify the overall coarsening kinetics. For an isotropic interfacial microstructure with an initial grain size $\bar{r}_i$, the mean field governing equation for the final grain size $r_{f(\xi)}$ is
\[
\frac{At}{a} = \int_{\bar{r}_i}^{\bar{r}_{f(\xi)}} \frac{d\bar{r}} {\left(e\,^{\xi/\bar{r}}- e^{(\xi-1)/\bar{r}}\right)}.
\]
Putting $y=1/\bar{r}$, with $Y_{f(\xi)}=(1/\bar{r}_{f(\xi)})$ and $Y_i=(1/\bar{r}_i)$,
\begin{align}
\label{eq:GGIntegralChangeOfVariable}
\frac{At}{a}=-\int_{Y_i}^{Y_{f(\xi)}} \frac{y\,e^{(1-\xi) y}}{e^{y}-1}\,\frac{dy}{y^3}.
\end{align}
This form is convenient for the following standard Maclaurin expansion (Ref.~\cite{book:AbramowitzStegun:1965}, p. 804):
\[
\frac{y\,e^{\lambda y}}{e^y-1}=\sum_{m=0}^\infty B_m(\lambda)\,\frac{y^m}{m!}\;.
\]
Here, $B_m(\lambda)$ are the \emph{Bernoulli polynomials}. They are tabulated (Ref.~\cite{book:AbramowitzStegun:1965}, p. 809); for example,
\begin{align}
B_0(\lambda)=1,\;  B_1(\lambda)=-\tfrac{1}{2}+\lambda\;\mbox{and}\;B_2(\lambda)=\tfrac{1}{6}-\lambda+\lambda^2\;.\nonumber
\end{align}
Also, $B_m(1-\lambda)=(-1)^mB_m(\lambda)$.
Thus, the integral in Eq.~\ref{eq:GGIntegralChangeOfVariable} can be written as
\[
\sum_{m=0}^\infty \frac{(-1)^mB_m(\xi)}{m!}\int_{Y_{f(\xi)}}^{Y_i} y^{m-3}\,dy\;,
\]
and the grain growth equation becomes
\begin{widetext}
\[
\frac{At}{a}=\left [\frac{1}{2y^2} - \frac{B_1(\xi)}{y} + \frac{B_2(\xi)}{2}\,\ln{y}-\sum_{m=3}^\infty C_m(\xi)\,y^{m-2}\right ]_{Y_i}^{Y_{f(\xi)}},\;\textrm{with}\;C_m(\xi)=\frac{(-1)^m B_m(\xi)}{(m-2)m!}\;.
\]
\end{widetext}
Remembering $\bar{r}$=$(1/y)$, we arrive at the coarsening law
\begin{align}
\label{eq:GGEquationExactSolution}
\frac{2At}{a} = & \,h\left[\bar{r}_{f(\xi)}\right] - h\left[\bar{r}_i\right]\;,\textrm{with}\\
h(\bar{r}) = & \,\bar{r}^2 - 2B_1(\xi)\bar{r} + B_2(\xi)\,\ln{\bar{r}}-2 \sum_{m=3}^\infty \frac{C_m(\xi)}{\bar{r}^{m-2}}.\nonumber
\end{align}
\begin{figure} [h!tb]
\includegraphics[width=\columnwidth]{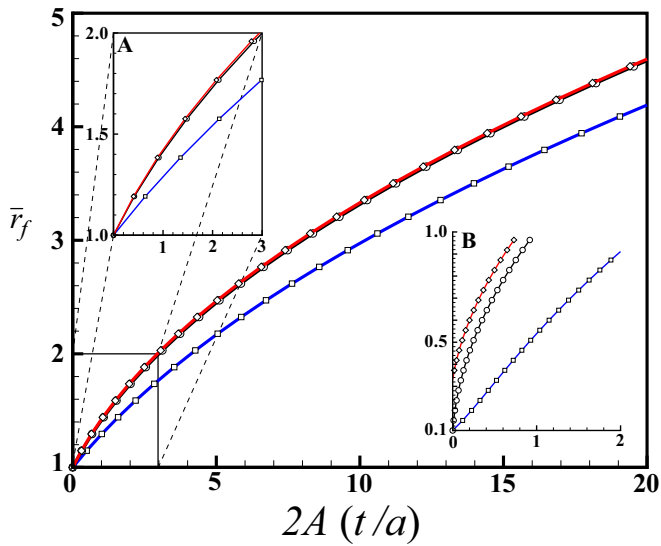}
\caption{\label{fig:subsuper-linearGGKinetics} (color online). Reduced grain size $\bar{r}_f$ vs. normalized time $2A(t/a)$ predicted by Eq.~\ref{eq:GGEquationExactSolution} for initial grain size $\bar{r}_i=1$. Circles, squares and diamonds represent linear, sub-linear ($\xi = 0$) and super-linear ($\xi = 0.5$) IM rates-based coarsening, respectively. Inset $A$ shows an enlarged view for $1 < \bar{r}_f  < 2$. Inset $B$ shows the variation for $\bar{r}_i = 0.1$ and $0.1 < \bar{r}_f < 1$.}
\end{figure}

At high temperatures and large grain sizes $\bar{r}\gg1$, Eq.~\ref{eq:GGEquationComplete} can be linearized and we recover a linear relation between IM rates and the driving force, $d\bar{r}/dt \approx A/a\bar{r}$. The resultant governing equation for coarsening kinetics is, 
\begin{align}
\label{eq:parabolicGrowthLaw}
\frac{2At}{a} = \int_{\bar{r}_i}^{\bar{r}_f(p)}\bar{r} d\bar{r} = \bar{r}_{f(p)}^2 - \bar{r}_i^2,
\end{align}
where $\bar{r}_{f(p)}$ is the final grain size. As expected, when $\bar{r}_{f(p)}\gg\bar{r}_i$,  $\bar{r}_{f(p)} \propto \sqrt{t}$. Combining Eqs.~\ref{eq:GGEquationExactSolution} and \ref{eq:parabolicGrowthLaw} yields the difference in grain size due to non-linear IM rates,
\begin{align}
&\bar{r}_{f(0)}^2 - \bar{r}_{f(p)}^2 \approx - \left\{ \left[\bar{r}_{f(0)}-\bar{r}_i \right] + \frac{1}{6}\ln\left[\frac{\bar{r}_{f(0)}}{\bar{r}_i}\right]\right\},\;\mbox{and}\nonumber\\
&\bar{r}_{f(0.5)}^2 - \bar{r}_{f(p)}^2 \approx \frac{1}{12}\ln\left[\frac{\bar{r}_{f(0.5)}}{\bar{r}_i}\right]\nonumber.
\end{align}

Coarsening kinetics described by Eq.~\ref{eq:GGEquationExactSolution} is shown in Fig.~\ref{fig:subsuper-linearGGKinetics}, a plot of $\bar{r}_{f(0)}$ and  $\bar{r}_{f(0.5)}$ against $2A(t/a) = h[\bar{r}_{f(\xi)}] - h[1]$, for $\bar{r}_i = 1$. Contribution of $O(\bar{r}^{-2})$ and higher terms is negligible and ignored. Grain size predicted by parabolic coarsening kinetics due to linear IM rates, $\bar{r}_{f(p)}$, is also shown for comparison. For sub-linear IM rates ($\xi = 0$), the linear term dominates during early stage coarsening. The negative deviation from classical parabolic coarsening kinetics increases linearly with grain size. We approach parabolic coarsening kinetics for $\bar{r}_{f(0)}\gg1$, yet the final grain size is smaller compared to classical parabolic coarsening - the coarsening is suppressed.
The effect is exaggerated at smaller initial grain sizes (inset $B$ in Fig.~\ref{fig:subsuper-linearGGKinetics}); coarsening kinetics is increasingly linear  ($\bar{r}_{f(0)} \le 1$).
For super-linear IM rates ($\xi = 0.5$),  the positive deviation from parabolic coarsening increases logarithmically and is much slower; the increase is substantial only for decades increase in grain sizes. The deviation is enhanced for $\bar{r}_i<1$, as IM rates are super-linear for larger range of grain sizes. This can be seen in inset $B$ in Fig. \ref{fig:subsuper-linearGGKinetics}, for $\bar{r}_i=0.1$.

Our analysis shows that a fundamental understanding of overdriven interface motion is critical for predicting  coarsening kinetics in nanocrystalline interfacial microstructures. Recent molecular dynamics (MD) simulations of flat bicrystals in pure Al have been performed at small and large driving forces, offering a basis for understanding the coarsening kinetics of grain boundary microstructures. The mobility of a  flat $\theta=38.2^\circ$ $<$$111$$>$ tilt misorientation grain boundary has been extracted in the zero driving force limit, and also under the influence of a bulk body force. Both studies were performed using an embedded-atom-method (EAM) framework for the inter-atomic potentials, justifying the comparison. The former is based on the random walk of the mean grain boundary position due to the uncorrelated thermal noise in the system~\cite{gbm:TrauttUpmanyuKarma:2006} and was extracted at a temperature $T=750^\circ$K, the latter on a synthetic driving force due to an orientation dependent bulk energy term at $T=800^\circ$K~\cite{gbm:JanssensHolm:2006}. The mobility of the synthetically driven grain boundary was extracted using a driving force $0.025$\,eV/atom. Using a conservative estimate of the activation volume, $\omega_m=\Omega_{Al}$, the driving force is still well past our critical value, $\beta p\omega_m=0.4$. Furthermore, the driving force for this boundary is of the order of the activation energy for migration of this boundary ($\sim0.02$\,eV), extracted in the zero driving force limit [ZTT and MU, to be published].

Comparison of the absolute mobilities grain boundaries extracted using these two techniques reveals that zero driving force limit mobility is faster by almost an order of magnitude ($4.4\times10^{-7}$ vs $6.0\times10^{-8}$ m$^4$J$^{-1}$s$^{-1}$)]), even though it was extracted at a slightly lower temperature. While systematic studies are necessary to understand the rather large difference (changes in energy of the activated state or large activation volume for migration), the decrease in mobility and therefore the IM rates due to non-linearities suggests that coarsening of nanocrystalline is intrinsically suppressed, also confirmed in experiments. This intrinsic effect must be factored in before ascribing the slow coarsening rates of these microstructures to extrinsic effects on interface motion. Meso-scale grain growth simulations are necessary to understand the combined effect.

MU and ADR acknowledge support from DOE-sponsored Computational Materials Science Network (CMSN) on ``{\it Dynamics and Cohesion of Materials Interfaces and Confined Phases Under Stress}'', and Office of Naval Research, Award N00014-06-1-0207 titled ``{\it Particle Strengthened Interfaces}". ADR also acknowledges support under the MRSEC program of the National Science Foundation under Award Number DMR-0079996.

\end{document}